\documentclass{article}
\usepackage[T1]{fontenc} 
\usepackage[utf8]{inputenc} 
\usepackage{ismir,amsmath,cite,url}
\usepackage{graphicx}
\usepackage{color}

\usepackage{subcaption} 

\usepackage{lineno}

\title{End-to-end adversarial white box attacks on music instrument classification}

\oneauthor
 {Katharina Prinz and Arthur Flexer}
 {Institute of Computational Perception, Johannes Kepler University Linz, Austria \\ {\tt \{katharina.prinz,arthur.flexer\}@jku.at}}

\DeclareMathOperator{\sign}{sign}
\DeclareMathOperator{\clip}{clip}

\begin{document}

\maketitle
\begin{abstract}
Small adversarial perturbations of input data are able to drastically change performance of machine learning systems, thereby challenging the validity of such systems.
We present the very first end-to-end adversarial attacks on a music instrument classification system allowing to add perturbations directly to audio waveforms instead of spectrograms. Our attacks are able to reduce the accuracy close to a random baseline while at the same time keeping perturbations almost imperceptible and producing misclassifications to any desired instrument.
\end{abstract}
\section{Introduction}
\label{sec:introduction}

Adversarial attacks add marginal and often imperceptible perturbations to input data, thereby drastically changing the performance of state of the art machine learning systems. This phenomenon has been first discovered in image object recognition\cite{Szegedy2014IntruigingProperties}, where high performing systems based on deep learning are highly sensible to imperceptible perturbations of images created by an adversary trying to fool the system.

While adversarial attacks typically do not pose a security threat in the field of Music Information Retrieval (MIR), they allow to question and analyse the validity of MIR systems, as is being done for a music instrument classification system in this paper. For this purpose we evaluate four different adversarial attack approaches on music instrument classification.
Starting from an audio waveform, the attacks produce perturbed versions, i.e. adversarial examples, that are misclassified as any instrument we choose, while having a very high signal to noise ratio (SNR) between original signal and added perturbation of up to 44.23dB on average. Contrary to previous results\cite{Kereliuk2015AudioAdversaries} only using spectral representations of audio as image inputs to algorithms from image recognition~\cite{Szegedy2014IntruigingProperties}, we also adapt adversarial attacks from speech recognition~\cite{Carlini2018STT}. This enables us to perform attacks in an end-to-end fashion, instead of on the spectrogram. Another novel aspect of our work is that our data are monophonic instrument or singing voice sounds, not polyphonic songs from different genres, which makes adversarial perturbations easier to hear and hence harder to hide in the original waveform.

In what follows, we start with reviewing related work in section~\ref{sec:relwork}, before we introduce our data (section~\ref{sec:data}) and instrument classification system as well as four end-to-end adversarial attacks in section~\ref{sec:methods}. Thereafter, we compare the different approaches in our experiments (section~\ref{sec:experiments}), and discuss and summarize results in sections~\ref{sec:evaluation_discussion} and \ref{sec:conclusion}.

\section{Related Work}
\label{sec:relwork}
Since the vulnerability of neural networks to adversarial examples was discovered~\cite{Szegedy2014IntruigingProperties}, numerous different approaches of adversarial attacks have been proposed. These attacks are often differentiated by how much knowledge they use, i.e.\ white box or black box attacks, and whether they aim to change the classification of an example to an arbitrary (untargeted attack) or specified new class (targeted attack). White box attacks are denoted as such whenever we assume (near) perfect knowledge about a system, and e.g. have access to a model and its parameters. For black box attacks we have very limited knowledge of a system and no access to elements such as model parameters.

A prominent early example of white box untargeted adversarial attacks is the Fast Gradient Sign Method (FGSM)~\cite{Goodfellow2015FGSM}. Instead of trying to decrease the loss of an image with respect to its ground-truth class as is done during training of a classifier, FGSM \emph{increases} the loss by pushing the image one step into the gradient direction of said loss function.
A second untargeted white box method is Projected Gradient Descent applied to the negative loss function (PGDn)~\cite{Madry2017PGD}, where the loss of an input example is maximised iteratively.

Further examples of adversarial attacks from the image domain, including both white and black box attacks, as well as untargeted and targeted methods, are Deepfool~\cite{Moosavi2016Deepfool}, the universal adversarial attack~\cite{Moosavi2017Universal}, the $L_0$, $L_2$ and $L_\infty$ attack~\cite{Carlini2017TowardsRobustness} and the one-pixel attack~\cite{Su2019OnePixel}.

In the field of speech recognition, a white box targeted attack has been proposed by Carlini \& Wagner~\cite{Carlini2018STT}, in which the computation of adversarial examples for an end-to-end speech-to-text system is formulated as an optimisation problem. This approach is particularly interesting due to its usage of audio and we will refer to it as C\&W in the remainder of our paper. Other approaches in the audio domain use psychoacoustic hiding to make adversarial perturbations as imperceptible as possible \cite{Schonherr2018PsychoacousticHiding,Qin2019ImperceptibleSpeech}. 

In work related to ours in this paper, an adversarial attack from the image domain (cf.\ \cite{Carlini2017TowardsRobustness}) is applied to various sound event classifiers with a particular focus on the transferability of adversarial examples between networks~\cite{Subramanian2020Study}. While Figure 1 in section 2 in ~\cite{Subramanian2020Study} suggests that the attack is performed end-to-end, this is not clarified any further in the remainder of the work; further work suggests, that there is still a lack of reliable attacks for raw waveform signals \cite{Esmaeilpour2020SecuringAudioClassification}. Different research directions in audio are the robustness of adversarial attacks in sound event classification \cite{Subramanian2019RobustnessOfAdversaries}, and increasing the robustness of audio classifiers against adversaries, e.g. by detecting adversarial examples \cite{Esmaeilpour2020DetectionSubspace,Esmaeilpour2020SecuringAudioClassification,Yang2019DetectionTemporalDependency}.

For the field of MIR, so-called \emph{irrelevant} random but linear time-invariant audio filtering transformations of the music signal have been used in an untargeted black box attack to both drastically deflate and inflate the performance~\cite{Sturm2014DetermineHorse} of genre classification systems. 
The irrelevance is ascertained with listening tests and real human subjects, with the transformation being audible but not changing the clear impression of a certain musical genre.
The same problematic behaviour has also been documented concerning music emotion classification~\cite{Sturm2014DetermineHorse} (again using filtering transformations) and rhythm classification \cite{Sturm2016Horse}, where a slight untargeted change in tempo is able to fool the classifier. 

Going beyond untargeted filtering approaches, first targeted white box attacks on genre recognition systems have been reported\cite{Kereliuk2015AudioAdversaries}, building upon approaches from image object recognition~\cite{Szegedy2014IntruigingProperties}. Essentially magnitude spectral frames computed from audio are treated as images, which however requires non-trivial care to ensure that produced adversarial examples are also valid audio signals. 

As for the MIR system we are trying to attack in this paper, instrument classification can be divided~\cite{Lostanlen2018Extended} into the following tasks, with increasing degree of difficulty: isolated note instrument classification, solo instrument classification, and multi-label classification in polyphonic mixtures. 
The data used in this paper is a subset derived from a recent audio tagging challenge (see section~\ref{sec:data}), essentially yielding isolated note and solo instrument data as well as singing voice audio. Our data is therefore an expanded version of isolated note classification, which has largely been solved (see e.g.\ \cite{Bhalke2015AutomaticMI}) and therefore is a good testbed for our adversarial attacks by providing unambiguous class information, clear spectral characteristics per instrument and a difficult setting for hiding adversarial information.


\section{Data}
\label{sec:data}
The data used in this work is a subset of the curated training set provided for Task 2 in the DCASE2019 Challenge~\cite{Fonseca2019Data}. More precisely, out of originally 4970 samples with diverse audio labels, we use all 799 mono audio files which are annotated with one single out of 12 different musical labels: accordion, acoustic guitar, bass drum, bass guitar, electric guitar, female singing, glockenspiel, gong, harmonica, hi-hat, male singing, and marimba / xylophone. The labels are similarly distributed, and occur at least 47 and at most 75 times among the 799 files. In what follows, we use the term \emph{instrument} to describe both instrumental and singing voice sounds.  

All audio files and their manually obtained labels originate from the Freesound Dataset~\cite{Fonseca2017Freesound}. The labels themselves are weak, i.e.\ only available at clip-level without starting and ending time information. Audio samples essentially consist of isolated notes or single instrument sounds as well as singing voice audio, with very few of the curated train samples having potentially incomplete labels. In other words, labelled instruments are always present in the audio, but there might be additional instrument sounds which are not represented in the respective annotation. 

To reduce computational expenses without significant loss of quality, we re-sample all audio files from 44.1 kHz to 16 kHz. The length of different audio samples varies from 0.3 to 29.29 seconds.


\section{Methods}
\label{sec:methods}

\subsection{Instrument classification system}
\label{subsec:instrument_recognition}

\textbf{Data Pre-Processing:} In contrast to previous work on adversaries in MIR, cf.\ \cite{Kereliuk2015AudioAdversaries}, in which adversarial examples are computed for spectrograms, we perform adversarial attacks end-to-end, i.e.\ on raw waveforms. This avoids having to restrict spectrograms in non-trivial ways to remain valid time-domain signals. To achieve this, we use a differentiable pre-processing module in front of the convolutional neural network (CNN) used for classification. By using torchaudio\footnote{https://pytorch.org/audio} to realise this first module, we can transform audio into a desired time-frequency representation, and still propagate gradients back to the original waveforms. 

To transform raw audio into the time-frequency domain, we first compute spectrograms with a Fast Fourier-Transform (FFT) size of 2048, an equally sized Hann-window and a hop size of 512. The spectrograms are subsequently mapped to Mel-scale, with 100 Mel-bands between 40 Hz and 8 kHz. Then, the resulting Mel-spectrograms are transformed to decibel (dB) scale and normalised to have zero mean and a standard deviation of one over the different Mel-bands, across all training data. Whenever we need to lengthen a spectrogram because of network requirements, we perform padding by repeating the spectrogram and inserting the repetitions with equal lengths both before and after the original time-frequency representation.

\begin{figure}
 \centerline{
 \includegraphics[width=\columnwidth]{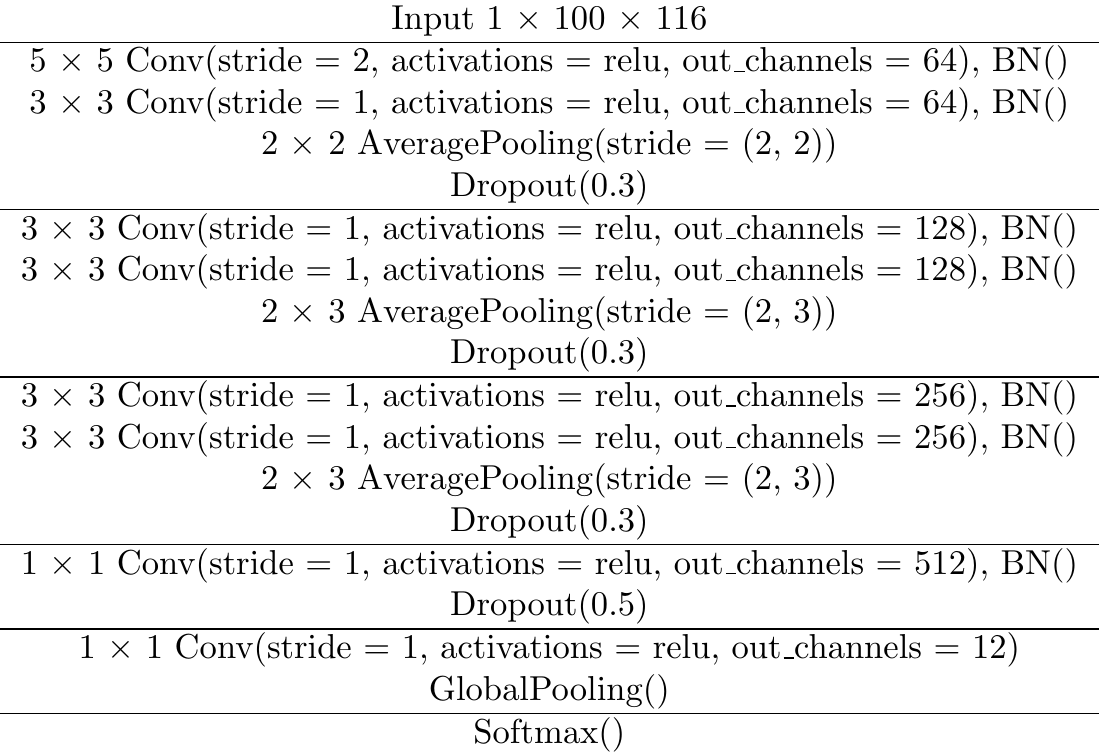}}
 \caption{Architecture of CNN used for instrument classification. $BN$ corresponds to Batch-normalisation layers; $GlobalPooling$ describes a global average-pooling over the frequency-, and a global max-pooling over the time-dimension. Input: $\text{channels} \times \text{Mel bands} \times \text{windows}$.}
 \label{fig:cnn}
\end{figure}

\noindent
\textbf{The Architecture:} In order to perform instrument classification, we use a CNN with its architecture depicted in \figref{fig:cnn}. The network consists of multiple convolutional layers with $5 \times 5$ and $3 \times 3$ convolutions and ReLU activations, followed by batch normalisation \cite{Ioffe2015BatchNorm} and average-pooling layers. Padding values correspond to the strides of convolutional layers. For regularisation, dropout is used. The final layers additionally contain $1 \times 1$ convolutions and a global pooling layer, which performs global average-pooling over the frequency-, and global max-pooling over the time-dimension of an input. 

\noindent
\textbf{Training and Evaluation Routine:} We split the data into a training and validation set containing 599 and 200 audio files respectively. Due to the small size of the dataset, we perform the evaluation of subsequent experiments on validation data.
To train the network, we use cross-entropy loss and the Adam optimiser with an initial learning rate of 0.001 for 150 epochs and a batch-size of 16. In addition, we use learning rate decay after 90 epochs with a multiplicative factor of 0.1. For normalisation of both training and validation data we use the global mean and standard deviation of all 100 different Mel-bands over the training data. 

To allow batch-wise training, we use windowed spectrograms of length 116. For shorter spectrograms, we perform padding as described above; for longer spectrograms, on the other hand, we extract half-overlapping windows of length 116 and choose an arbitrary one for each iteration. When performing validation, we use spectrograms with their original lengths, or padded versions to fulfil minimum length requirements of the network.

\subsection{Adversarial Attacks}
\label{subsec:attacks}
In our experiments, we apply four different adversarial white box attacks to an instrument classification system. Two of these attacks, namely FGSM and PGDn (cf. section \ref{sec:relwork}), are untargeted. The other two are targeted attacks, and both adaptations of the C\&W method (cf. section \ref{sec:relwork}). The first version is a slightly modified C\&W method, adapted to be suitable for our system. In the second targeted attack, we incorporate a new loss function modelling audio domain knowledge.  

To clarify the differences of these approaches, we can look at their optimisation goals which define how an adversarial perturbation is computed. In what follows, $\nabla$ corresponds to a gradient and $L_\mathrm{net}$ to the cross-entropy loss function, which we also use to train our model. Note that $f$ describes our instrument classification system, including both the pre-processing module as well as the classifier, and $x$ and $y$ denote the original signal and its ground-truth label. The adversarial perturbation that we want to compute is $\delta$, i.e.\ an adversarial example can be expressed by $x + \delta$. The perturbation $\delta$ is always initialised with zeros unless stated otherwise.  For iterative methods, $\eta$ describes the multiplication factor for gradient updates, and $k$ corresponds to the current iteration. Target predictions for targeted attacks are denoted by $t$. Furthermore, we use a factor $\alpha$ for the targeted methods to control the trade-off between finding adversarial perturbations more easily, and keeping perturbations as small as possible. All gradient updates are performed based on the sign of the gradient.

\noindent
\textbf{FGSM:} The Fast Gradient Sign method~\cite{Goodfellow2015FGSM} is the only single step approach, for which we compute the adversarial perturbation $\delta$ by moving a scaled step in the gradient direction in order to increase the loss, i.e.
\begin{equation}\label{eq:fgsm}
  \delta = \lambda * \sign(\nabla_x \ L_\mathrm{net}(f(x), y)).
\end{equation}
The parameter $\lambda$ is a scaling factor controlling the trade-off between difficulty of finding adversarial examples and the quality thereof.

\noindent
\textbf{PGDn:} Projected Gradient Descent on the negative loss function~\cite{Madry2017PGD} is an untargeted attack with the purpose of increasing the loss of a sample iteratively. For this attack, we start at a random point close to the original signal, i.e. we initialise the adversarial perturbation by sampling from a uniform distribution $\mathcal{U}(-\epsilon, \epsilon)$. After initialisation, we update the perturbation by computing
\begin{equation}\label{eq:pgdn}
  \delta_{k+1} = \clip_\epsilon(\delta_k + \eta * \sign(\nabla_{\delta_k} \ L_\mathrm{net}(f(x + \delta_k), y))).
\end{equation}
In each iteration, $\delta$ is clipped to stay between $-\epsilon$ and $\epsilon$.

\noindent
\textbf{C\&W:} In contrast to the untargeted attacks, in which gradient \emph{ascent} is used to increase the loss on a ground-truth label to change the prediction of a classifier, we use gradient \emph{descent} for the targeted methods to minimise the loss with respect to a particular target class $t$. Instead of minimising the network loss directly, the C\&W method optimises a weighted combination of squared L2-norm of the perturbation and a so-called CTC-loss function~\cite{Carlini2018STT}. As the CTC-loss function is specific to speech-to-text systems, we substitute it with the cross-entropy loss function. Again, C\&W clips the perturbation $\delta$ in each iteration, leading to
\begin{align}\label{eq:cw}
  L_\mathrm{total} &= \|\delta_k\|^2_2 + \alpha * L_\mathrm{net}(f(x + \delta_k), t), \nonumber \\
  \delta_{k+1} &= \clip_\epsilon(\delta_k - \eta * \sign(\nabla_{\delta_k} L_\mathrm{total})).
\end{align}
The original C\&W method uses a re-scaling factor to scale clipped adversarial perturbations in order to stay below certain SNR-thresholds~\cite{Carlini2018STT}. This factor is decreased as soon as an initial adversarial example is discovered; we omit the re-scaling factor in \eqnref{eq:cw}, as we stop attacks immediately after an initial perturbation is found.

\noindent
\textbf{Multi-Scale C\&W:} The fourth adversarial attack is a different adaptation of C\&W~\cite{Carlini2018STT}. As opposed to using the squared norm to keep perturbations small, we use a multi-scale loss~\cite{Engel2020DDSP} in order to make original and adversarial waveform sound as similar as possible. More precisely, we use
\begin{align}\label{eq:ms_cw}
  L_\mathrm{spec} &= \sum_i \| X_i - \hat{X}_i \|_1 + \| \log X_i - \log\hat{X}_i \|_1 \nonumber, \\
  L_\mathrm{total} &= L_\mathrm{spec} + \alpha * L_\mathrm{net}(f(x + \delta_k), t), \nonumber \\
  \delta_{k+1} &= \clip_\epsilon(\delta_k - \eta * \sign(\nabla_{\delta_k} L_\mathrm{total})).
\end{align}
Note that $\hat{x}$ is short for an adversarial example, i.e. $x + \delta_k$, and upper-case letters describe a signal transformed to the time-frequency domain. The index $i$ is used to iterate over spectra resulting from different Fourier transformation sizes (2048, 1024, 512, 256, 128, 64; cf. \cite{Engel2020DDSP}).
\section{Experiments}
\label{sec:experiments}

\subsection{Experimental Setup}
\label{subsec:setting}
In what follows, we compare the four end-to-end adversarial attacks introduced in section~\ref{subsec:attacks}. For three of the four methods we use Adam \cite{Kingma2015Adam} to perform gradient descent and ascent respectively. As FGSM requires only a single gradient update, we apply stochastic gradient ascent instead of using Adam. We also apply a grid-search to find attack-specific parameters, and restrict each iterative method to find an adversarial perturbation within 500 iterations. To further reduce complexity, we sample random target classes different from a sample's original prediction once for targeted attacks, and use these targets in all parameter settings within the grid-search. More information on used parameters for the different methods can be found in section A.1 in the supplementary material. To allow reproducibility of results, all code to run the experiments is available via github\footnote{https://github.com/msBluemchen/adversarial\_instrument\_classification}.

\subsection{Results}
\label{subsec:results}
In~\tabref{tab:results}, we compare the four adversarial attacks introduced in section~\ref{subsec:attacks} based on multiple factors. First, we state the number of adversarial examples found for our 200 validation samples, followed by the average accuracy on the validation set with respect to the ground-truth labels. This accuracy is computed on the adversarial signal of a sample if one was found, or else on the original signal. The last three columns are computed solely on adversarial examples, and depict the average SNR, the median of iterations it takes an adversary to find an adversarial perturbation, and finally the average confidence of our system when classifying an adversarial example. The confidence of a sample here refers to the probability that our system assigns to its predicted class. As the initial perturbation for PGDn and the targets for C\&W and Multi-Scale C\&W are sampled randomly, we repeat these experiments five times and state mean and standard deviation over all runs. Lines 3 to 6 in~\tabref{tab:results} are based on the parameter settings in our grid-search that achieve the highest average SNR with at least 150 found adversarial examples. In lines 7 to 10, we show how these results change when requiring at least 180 found adversarial examples. Whenever we refer to differences being \emph{statistically significant} in the remainder of this section, it means significant as tested via paired t-tests at a 5\% error level; for more detailed results, including exact t-test results, we refer to section A.1 in the supplementary material.

\begin{table*}[ht]
	\begin{center}
		\begin{tabular}{| l | c | c | r | r | r |}
			\hline
			Data Origin & \# Samples & Accuracy & SNR & Iterations & Confidence\\
			\hline
			\hline
			clean & 200 & 0.835 & - & - & $0.93$ \\ \hline
			\hline
			FGSM & 153 & 0.250 & $-7.74$ & $1.0$ & $0.72$ \\ 
			PGDn & $151.8 \pm 0.7$* & $0.171 \pm 0.004$* & $40.13 \pm 0.05$* & $15.8 \pm 0.4$* & $0.55 \pm 0.00$* \\ \hline 
			C\&W & $153.2 \pm 2.6$* & $0.201 \pm 0.016$* & $44.23 \pm 0.37$* & $51.4 \pm 2.7$* & $0.45 \pm 0.00$* \\ 
			$\text{C\&W}_{multi\_scale}$ & $163.6 \pm 3.0$* & $0.167 \pm 0.012$ * & $43.82 \pm 0.09$* & $71.6 \pm 5.4$* & $0.44 \pm 0.00$* \\ \hline \hline 
			FGSM & 179 & 0.130 & -24.83 & 1.0 & 0.86 \\ 
			PGDn & $190.8 \pm 1.2$* & $0.026 \pm 0.004$* & $16.47 \pm 0.10$* & $2.0 \pm 0.0$* & $0.74 \pm 0.01$* \\ \hline 
			C\&W & $180.2 \pm 2.3$* & $0.094 \pm 0.010$* & $42.98 \pm 0.18$* & $66.1 \pm 3.7$* & $0.44 \pm 0.01$* \\ 
			$\text{C\&W}_{multi\_scale}$ & $196.4 \pm 1.0$* & $0.024 \pm 0.004$* & $39.49 \pm 0.17$* & $22.6 \pm 1.0$* & $0.49 \pm 0.01$* \\ \hline 
		\end{tabular}
	\end{center}
	\caption{Comparison of four end-to-end adversarial attacks on our instrument classification system. Results are chosen based on largest SNR with at least 150 (lines 3 to 6) and 180 (lines 7 to 10) out of 200 adversarial examples respectively. Depicted are averages or the median over samples; for PGDn, C\&W and Multi-Scale C\&W additionally average and standard deviation* of results over 5 runs are stated.}
	\label{tab:results}
\end{table*}

\noindent
\textbf{Number of Samples:} The results for each of the four adversarial attacks are chosen in a way, that either at least 150 or 180 adversarial examples are found. The overall lowest number is achieved by PGDn, with an average of 151.8. Both when requiring at least 150 or 180 samples, Multi-Scale C\&W is the method that finds the highest amount of adversarial examples, with on average 163.6 and 196.4 samples. The difference to the remaining methods is statistically significant. Note, that for all of the four methods except of FGSM it is possible to find at least 180 adversarial examples; for the single-step method, the highest amount of samples that is found within our grid-search is 179. 

\noindent
\textbf{Accuracy:} The average accuracy on validation data is strongly influenced by the number of adversarial examples an adversary finds. The more often an adversary is successful in finding adversarial perturbations, the lower the accuracy tends to go; exceptions are adversarial examples that lead to correct predictions for previously misclassified samples. After an adversarial attack, the reduced accuracy of our system is always close to the baseline accuracy of 0.125, which can be achieved by predicting only the most frequent class within the validation set ('Gong'). More precisely, the average accuracy is between 0.250 and 0.167 when we require 150 adversarial examples found by an adversary, and by requiring at least 180 examples, this even drops to values between 0.130 and 0.024. In both scenarios, FGSM achieves the highest accuracy with 0.250 and 0.130, followed by the targeted C\&W method with averages of 0.201 and 0.094. PGDn leads to the second lowest accuracies with 0.171 and 0.026. Multi-Scale C\&W results in the lowest average accuracy of 0.167 and 0.024, which is statistically significant compared to FGSM and C\&W. 

To put these results into perspective, we take a look at how the accuracy of our system changes if we add random white noise instead of adversarial perturbations to validation samples. When we overlay original signals with white noise of a similar average SNR as the two targeted attacks (42.16dB), the accuracy of our system remains with 0.79 close to the accuracy of 0.835 on clean data. 

\begin{figure*}
    \begin{subfigure}{.33\textwidth}
      \centering
      \includegraphics[trim={2.5cm 0.5cm 2.5cm 1cm},clip,height=.75\linewidth]{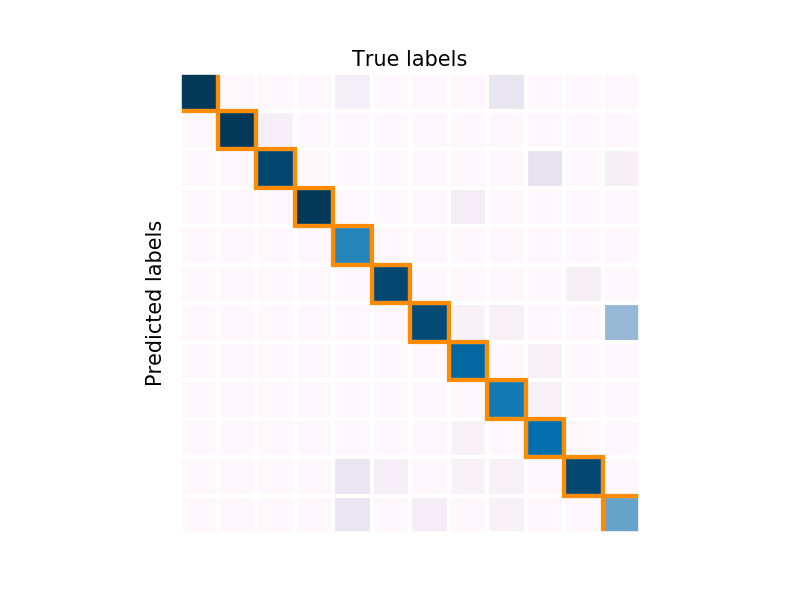}  
      \captionsetup{justification=centering}
      \caption{Clean data \\ \hspace{1cm}}
      \label{subfig:clean_mcm}
    \end{subfigure}%
    \begin{subfigure}{.33\textwidth}
      \centering
      \includegraphics[trim={2.5cm 0.5cm 2.5cm 1cm},clip,height=.75\linewidth]{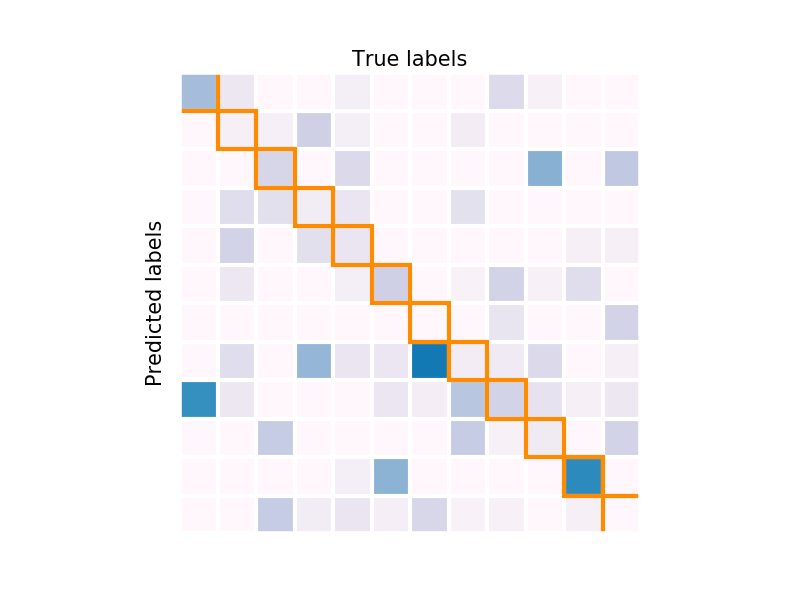}  
      \captionsetup{justification=centering}
      \caption{PGDn data \\ ($\epsilon = 0.0005$ and $\eta = 1e-05$)}
      \label{subfig:pgdn_mcm}
    \end{subfigure}%
    \begin{subfigure}{.33\textwidth}
      \centering
      \includegraphics[trim={2.5cm 0.5cm 2.5cm 1cm},clip,height=.75\linewidth]{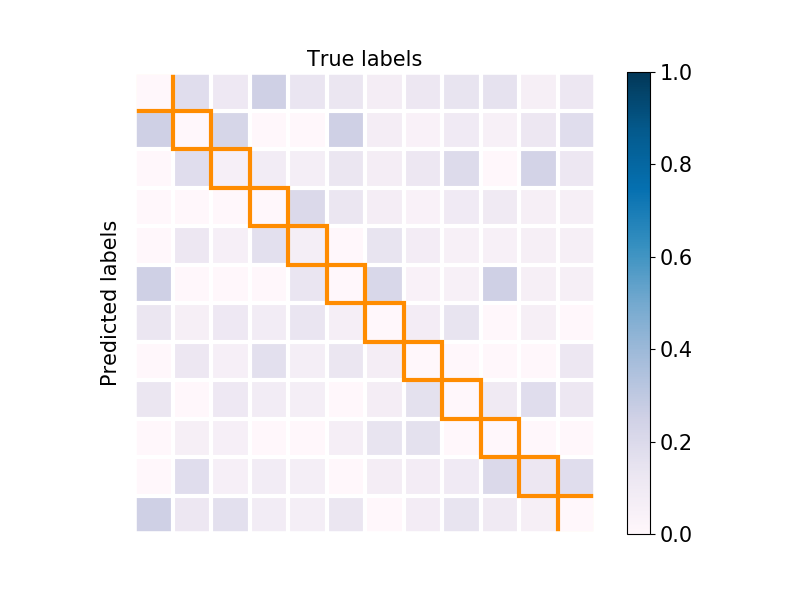}  
      \captionsetup{justification=centering}
      \caption{Multi-Scale C\&W data \\ ($\alpha = 15, \epsilon = 0.01, \eta = 5e-05$)}
      \label{subfig:mscw_cm}
    \end{subfigure}
    \caption{Confusion matrices computed on validation data, showing correct predictions in the diagonal, confusions off-diagonal. For samples without adversarial counterpart, original audio is used. Columns are ground-truth labels and rows predictions; in addition, columns are normalised to sum to 1. Order of labels (left to right and top to bottom): Accordion, Acoustic guitar, Bass drum, Bass guitar, Electric guitar, Female singing, Glockenspiel, Gong, Harmonica, Hi-hat, Male singing, Marimba and xylophone.}
\label{fig:conf_matrices}
\end{figure*}

In~\figref{fig:conf_matrices}, three confusion matrices illustrate the accuracy of our system after different attacks. The columns are ground-truth labels, and the rows represent predictions; correct predictions are in the diagonal, and confusions off-diagonal. Each column is normalised to sum to 1, such that a column shows the percentage of correct and incorrect classifications for one ground-truth class. For detailed confusion matrices including numerical values, we refer to section A.2 in the supplementary material. The leftmost image shows the performance on clean data. The majority of predictions is concentrated in the diagonal of the matrix; the classes which are least often classified correctly by our system are 'Marimba and xylophone' and 'Electric guitar'. The remaining two matrices are chosen to depict an untargeted as well as a targeted attack leading to a broad variety of new classifications.
The confusion matrix in the middle shows the confusions of our system after a PGDn attack. The untargeted attack leads to diverse new predictions, and appears to have the most difficulties with finding adversarial examples for samples with the ground-truth class 'Male singing', which results in the only noticeable diagonal entry  in~\figref{fig:conf_matrices}. When allowing bigger perturbations for the untargeted methods FGSM and PGDn (see sections A.2.2 and A.2.3 in the supplementary material), the confusions change significantly, as more and more samples with various ground-truth classes are classified as one of 'Gong', 'Harmonica' or 'Hi-hat'. Finally, the rightmost confusion matrix in~\figref{fig:conf_matrices} depicts confusions after a targeted Multi-Scale C\&W attack. The new predictions are very diverse, and values in the diagonal of the matrix are close or equal to 0.

\noindent
\textbf{SNR:} The signal to noise ratio (SNR) is computed as the ratio between original signal and added perturbation, with higher values signifying higher quality of audio, or less perceptible perturbations. Allowing bigger perturbations simplifies the task of finding higher numbers of adversarial examples, but it tends to decrease the average SNR. To clarify the level of perceptibility of different adversarial examples, we refer to the hearing examples in section A.3 in our supplementary material\footnote{https://msbluemchen.github.io/adversarial\_instrument\_classification}. Attacks with high SNR around 40dB can hardly be perceived at all or as low-volume high-frequency noise. All four attacks achieve higher average SNRs when we require at least 150 as opposed to 180 adversarial examples to be found. FGSM leads to the lowest SNRs with averages of -7.74dB and -24.83dB; while in the first case, this leads to clearly perceptible perturbations with still recognisable classes, the second case leads to distorted and therefore not meaningful adversarial examples. In order to find adversarial perturbations with FGSM that are less perceptible, significantly smaller scaling values $\epsilon$ are required, leading to small numbers of found adversarial examples. The average SNRs for PGDn, C\&W and Multi-Scale C\&W when requiring at least 150 adversarial examples are very similar, and for all three methods around 40dB. The highest average of 44.23dB is achieved with the C\&W method. 

When increasing the amount of required adversarial examples to be at least 180, the average SNRs for both C\&W and Multi-Scale C\&W decrease slightly to 42.98dB and 39.49dB respectively; the average SNR of PGDn, however, drops significantly to 16.47dB. This leads to perceptible adversarial perturbations of PGDn, while the original class remains recognisable. The difference of the SNR of C\&W compared to the remaining three methods is statistically significant. 

\noindent
\textbf{Iterations:} As the number of iterations it takes for an adversary to find a perturbation for a particular sample strongly varies, iterations are stated as the median or the average $\pm$ the standard deviation over medians of different runs. By definition, the single-step FGSM is the adversary requiring the least amount of iterations. The second least iterations are needed by PGDn, taking averages of 15.8 and 2.0 iterations. For the two targeted methods, the number of needed iterations appears to depend on specific parameter settings; in the first case, C\&W requires 51.4 iterations as opposed to 71.6 for Multi-Scale C\&W. In the second scenario however, Multi-Scale C\&W takes less iterations on average with 22.6 iterations in contrast to 66.1 by C\&W. 

In addition to looking at the number of iterations that different attacks take in various parameter settings, we compare runtime complexities of single iterations with respect to the waveform length $N$. The gradient $\nabla_{\delta} L_\mathrm{net}(f(x + \delta), y)$ is computed in each of the four methods, which is why we consider it to be a given and focus on the complexity of the remaining computations. For FGSM, PGDn and C\&W these computations are bounded by $N$, which is why the attacks are all $O(N)$. Multi-Scale C\&W on the other hand is dominated by the Fourier transformations in the computation of $L_\mathrm{spec}$ in~\eqnref{eq:ms_cw}, and is hence $O(N \log N)$.

\noindent
\textbf{Confidence:} The confidence of our system when classifying adversarial examples tends to decrease with increasing SNRs. More precisely, FGSM achieves the highest confidences with averages of 0.72 and 0.86. Nevertheless, the average confidence of the system does not fall below 0.44. 

\subsection{Using Targeted Attacks to Predict Accordion}
\label{subsec:targetattack}
Instead of randomly choosing a target class when using the two targeted end-to-end attacks, we can also try to transform any class into a particular target class. To demonstrate this, we use the 'Accordion' class, as it is one of the hardest target classes to reach in prior experiments. For C\&W, we find adversarial perturbations for all but 4 samples with the ground-truth classes 'Gong' and 'Male singing'. The adversarial examples have an average SNR of 37.41dB$ \pm 10.72$ and are predicted with an average confidence of $0.44 \pm 0.12$. As a median, the adversary takes 29 iterations to find a successful adversarial perturbation.

For Multi-Scale C\&W, the adversary is successful in finding adversarial examples for all but 7 samples. The adversarial perturbations lead to an average SNR of 37.57dB$ \pm 10.82$, and are classified with an average confidence of $0.44 \pm 0.11$. The median of required iterations is 33. For more information on the parameters we use for the two methods, we refer to section A.2.6 in the supplementary material.

\section{Discussion}
\label{sec:evaluation_discussion}
In summary, we see that all four adversarial attacks are able to significantly reduce the accuracy of our instrument classification system. The first untargeted attack, FGSM, is the least complex of the algorithms due to its single-step nature, but also leads to the largest perturbations. PGDn is slightly more complex, and leads to very similar predictions for all adversarial examples whenever high perturbations are allowed. The quality of audio is improved in comparison to the results of FGSM, as we can reach SNRs comparable to the targeted attacks.

The best attack based on the average SNR within our experiments is the targeted C\&W method. Despite its comparably high median of required iterations, it still has an asymptotically lower runtime complexity than the targeted Multi-Scale C\&W, and reaches average SNRs of above 40dB. The increase of the runtime complexity due to the computation of Fourier transforms within Multi-Scale C\&W does not always pay off in comparison to C\&W, particularly in cases where we want to find large numbers of adversarial examples. If however, different from our results in section~\ref{subsec:results}, no particular number of adversarial examples is required, Multi-Scale C\&W can be used to find perturbations with the overall highest average SNR; within our grid-search, the maximal average SNR for C\&W is 46.87dB (52 adversarial examples), while Multi-Scale C\&W reaches 64.26dB (3 adversarial examples). Both targeted attacks can be used to transform any class to a particular target prediction, as is demonstrated with the example of the `Accordion' class.

To put our results into perspective, we look at results from the one already published comparable adversarial attack in MIR~\cite{Kereliuk2015AudioAdversaries}. The approach allows setting confidence levels as well as minimal SNRs of adversarial examples. The authors achieve an average SNR of 23.0dB and 15.78dB respectively for two datasets, with a confidence level of 0.9. This setting allows them to reduce the accuracy of genre recognition systems from 0.490 to 0.117 and from 0.632 to 0.413. When decreasing the confidence level to 0.5 for the model with an initial accuracy of 0.632, an average SNR of 11.15dB and a new accuracy of 0.28 is obtained. In a different experiment, in which audio files are transformed to all existing classes, confidence levels of 0.5 and 0.9 respectively lead to average SNRs of 34.5dB and 26.8dB. These results are comparable to ours, as higher SNRs tend to lead to slightly lower average confidence levels. For average confidences between 0.44 and 0.55, we achieve average SNRs between 39.49dB and 44.23dB. Also, we decrease our clean accuracy of 0.835 to accuracies never higher as 0.201. Therefore, while our end-to-end methods do not support setting confidence level or SNR beforehand, performing a search over different parameter settings allows us to find examples with slightly higher average SNR and much more reduced accuracy.

The existence of adversarial examples suggests that "being able to correctly label the test data does not imply that our models truly understand the tasks we have asked them to perform”\cite{Goodfellow2015FGSM} and that impressive results of performance at almost human level might not use musical knowledge at all\cite{Sturm2013Classification,Sturm2014DetermineHorse}. This directly brings us to the question of validity, i.e.\ whether our instrument classification experiment is actually measuring what we intended to measure (see \cite{trochim2000research,urbano2013evaluation}). Our real intention is to produce algorithms classifying instrument sounds in general, not just for 799 instrument sounds in our data base, which of course extends the target population to all instrument sounds including those with perturbations. If such small perturbations, clearly not changing the perceived instrument characteristics, have such a large impact on the classification system, it must rely on a confounding factor exposing a problem of internal validity, with
no causal relation between type of instrument represented in the audio and instrument label
returned by the classifier. It is also a problem of external validity, with observed results
not generalizing from the original sound material to slightly changed versions thereof. After we have successfully exposed these validity and reliability problems in this paper, our future work will aim at explaining what the confounding factors are and, based on these insights, design a music instrument classification system that is robust against adversarial attacks.


\section{Conclusion}
\label{sec:conclusion}
In this work, we applied four different white-box adversarial attacks to an instrument classification system. Going beyond previous attacks, we computed adversarial examples directly for the waveform in an end-to-end fashion instead of starting from the spectrogram. The four adversaries were compared with respect to various factors such as runtime complexity, number of adversarial examples that can be found and perceptibility of the adversarial perturbations based on the SNR and listening examples. 
The successful adversarial attacks were able to decisively reduce accuracy of the classification system with adversarial perturbations of waveforms being almost imperceptible. This paper therefore is a contribution to the growing body of work that questions the validity of MIR results by exposing them to adversarial attacks, thereby gaining deeper understanding of their mode of operation.

\section{Acknowledgments}
\label{sec:acknowledgments}
This work is supported by the Austrian National Science Foundation (FWF, P 31988).

\bibliography{ms}

\end{document}